\documentclass{article}
\usepackage{amsmath}
\usepackage{cite}
\usepackage{graphicx}
\usepackage{dcolumn}

\newtheorem{conj}{Conjecture}

\begin{document}

\date{}
\title{Exact solutions, critical parameters and accidental degeneracy for the
hydrogen atom in a spherical box}
\author{Francisco M. Fern\'{a}ndez\thanks{%
fernande@quimica.unlp.edu.ar} \\
%EndAName
INIFTA, DQT, Sucursal 4, C. C. 16, \\
1900 La Plata, Argentina}
\maketitle

\begin{abstract}
We derive some properties of the hydrogen atom inside a box with an
impenetrable wall that have not been discussed before. Suitable scaling of
the Hamiltonian operator proves to be useful for the derivation of some
general properties of the eigenvalues. The radial part of the
Schr\"{o}dinger equation is conditionally solvable and the exact polynomial
solutions provide useful information. There are accidental degeneracies that
take place at particular values of the box radius, some of which can be
determined from the conditionally-solvable condition. Some of the roots
stemming from the conditionally-solvable condition appear to converge
towards the critical values of the model parameter. This analysis is
facilitated by the Rayleigh-Ritz method that provides accurate eigenvalues.
\end{abstract}

\section{Introduction}

\label{sec:intro}

Quantum mechanical models of particles confined within boxes of different
shapes have received considerable attention for many years\cite
{F01,SBC09a,SBC09b,S14}. In such reviews one can find all kind of atomic and
molecular systems enclosed inside surfaces that are impenetrable or
penetrable. In a recent paper, Amore and Fern\'{a}ndez\cite{AF25} came
across a most interesting accidental degeneracy that had not been discussed
before. The purpose of this paper is the analysis of possible accidental
degeneracies in the case of the hydrogen atom in a spherical box with the
nucleus clamped at origin.

In section~\ref{sec:model} we discuss the model and some of its mathematical
properties. In section~\ref{sec:CS} we investigate exact polynomial
solutions to the radial part of the Schr\"{o}dinger equation. In section~\ref
{sec:RR} we obtain accurate eigenvalues by means of the Rayleigh-Ritz method
(RRM)\cite{P68,SO96}. Finally, in section~\ref{sec:conclusions} we summarize
the main results of the paper and draw conclusions.

\section{The model}

\label{sec:model}

In this section we present the model and discuss some of the properties of
the time-independent Schr\"{o}dinger equation. We are interested in the
eigenvalue equation $H\psi =E\psi $ for the Hamiltonian operator
\begin{equation}
H=-\frac{\hbar ^{2}}{2m_{e}}\nabla ^{2}-\frac{K}{r},  \label{eq:H}
\end{equation}
where $m_{e}$ is the electron mass and $K>0$ is the strength of the Coulomb
potential (with units of energy$\times $length). For simplicity, we assume
that the nucleus is clamped at the origin. The solutions $\psi (r,\theta
,\phi )$ in spherical coordinates $x=r\sin \theta \cos \phi $, $y=r\sin
\theta \sin \phi $, $z=r\cos \theta $, $0<\theta <\pi $, $0\leq \phi <2\pi $%
, satisfy the boundary condition $\psi \left( r_{0},\theta ,\phi \right) =0$
because of the impenetrable wall of a spherical box of radius $r_{0}$.
Therefore, $0<r\leq r_{0}$.

In order to facilitate the mathematical treatment of the problem it is
convenient to carry out the scaling transformation $(x,y,z)\rightarrow (L%
\tilde{x},L\tilde{y},L\tilde{z})$, $r\rightarrow L\tilde{r}$, $\nabla
^{2}\rightarrow L^{-2}\tilde{\nabla}^{2}$, where $L$ is an arbitrary length,
that leads to\cite{F20}
\begin{equation}
H=\frac{\hbar ^{2}}{m_{e}L^{2}}\left( -\frac{1}{2}\tilde{\nabla}^{2}-\frac{%
m_{e}LK}{\hbar ^{2}\tilde{r}}\right) .  \label{eq:H_dim_L}
\end{equation}
The dimensionless box radius is $\tilde{r}_{0}=r_{0}/L$. If $E\left(
r_{0},K\right) $ denotes an eigenvalue of $H$, then this equation tells us
that
\begin{equation}
E\left( r_{0},K\right) =\frac{\hbar ^{2}}{m_{e}L^{2}}E\left( \frac{r_{0}}{L},%
\frac{m_{e}LK}{\hbar ^{2}}\right) .  \label{eq:E(r0,K)_gen}
\end{equation}
If $L=r_{0}$ we have
\begin{equation}
E\left( r_{0},K\right) =\frac{\hbar ^{2}}{m_{e}r_{0}{}^{2}}E\left( 1,\beta
\right) ,\;\beta =\frac{m_{e}r_{0}K}{\hbar ^{2}}.  \label{eq:E(r0,K)_r0}
\end{equation}
On the other hand, if $L=\hbar ^{2}/\left( m_{e}K\right) $ we have
\begin{equation}
E\left( r_{0},K\right) =\frac{\hbar ^{2}}{m_{e}r_{0}{}^{2}}\beta ^{2}E\left(
\beta ,1\right) .  \label{eq:E(r0,K)_L1}
\end{equation}
It follows from equations (\ref{eq:E(r0,K)_r0}) and (\ref{eq:E(r0,K)_L1})
that
\begin{equation}
E(1,\beta )=\beta ^{2}E(\beta ,1).  \label{eq:E(1,beta),E(beta,1)}
\end{equation}
It is clear that $E(1,\beta )$ is the dimensionless energy of a hydrogen
atom with interaction $-\beta /r$ and unit box radius, while $E(\beta ,1)$
is the dimensionless energy of a hydrogen atom with interaction $-1/r$ and a
dimensionless box radius equal to $\beta $. Both descriptions of the problem
are related by the simple expression (\ref{eq:E(1,beta),E(beta,1)}).

The Schr\"{o}dinger equation in any of the two cases discussed above is
separable in spherical coordinates as $\psi _{nlm}(r,\theta ,\phi
)=R_{nl}(r)Y_{l}^{m}(\theta ,\phi )$, where $n=0,1,\ldots $,  $l=0,1,\ldots $
and $m=0,\pm 1,\pm 2,\ldots ,\pm l$ are the radial, angular and magnetic
quantum numbers, respectively, and $Y_{l}^{m}$ are the well-known spherical
harmonics. The energy eigenvalues depend only on $n$ and $l$ so that we
write them as $E_{nl}$ from now on. For convenience, we do not resort to the
principal quantum number $n_{p}=n+l+1=1,2,\ldots $ that is mostly useful in
the case of the free hydrogen atom.

It is clear that $E_{nl}(1,0)$ are the eigenvalues of a free electron in a
box of unit radius; therefore, they are all positive. On the other hand,
\begin{equation}
\lim\limits_{\beta \rightarrow \infty }E_{nl}(\beta ,1)=\lim\limits_{\beta
\rightarrow \infty }\beta ^{-2}E_{nl}(1,\beta )=E_{nl}^{H}=-\frac{1}{%
2(n+l+1)^{2}},  \label{eq:E_HA}
\end{equation}
are the dimensionless energies of the free atom. In this case $%
E_{nl}^{H}>E_{n^{\prime }l^{\prime }}^{H}$ if $n+l>n^{\prime }+l^{\prime }$.
This obvious inequality will be useful later on.

Since $E_{nl}(1,0)=E_{nl}^{PB}>0$ and $\lim\limits_{\beta \rightarrow \infty
}\beta ^{-2}E_{nl}(1,\beta )<0$, then for each eigenvalue $E_{nl}$ there is
a value $\beta =\beta _{nl}^{c}$ such that $E_{nl}\left( \beta
_{nl}^{c},1\right) =E_{nl}\left( 1,\beta _{nl}^{c}\right) =0$. We will
calculate some of these critical values of $\beta $ in section~\ref{sec:RR}.

\section{Exact polynomial solutions}

\label{sec:CS}

The radial part of the Schr\"{o}dinger equation for the dimensionless
Hamiltonian operator
\begin{equation}
H=-\frac{1}{2}\nabla ^{2}-\frac{\beta }{r},  \label{eq:H_dim_beta}
\end{equation}
is
\begin{equation}
\mathcal{H}R(r)=ER(r),\;\mathcal{H}=-\frac{1}{2r^{2}}\frac{d}{dr}r^{2}\frac{d%
}{dr}+\frac{l(l+1)}{2r^{2}}-\frac{\beta }{r},  \label{eq:radial}
\end{equation}
with the boundary condition $R(1)=0$. This eigenvalue equation admits some
exact polynomial solutions because it is conditionally solvable (see, for
example, \cite{CDW00,T16} and references therein). In order to derive them
we propose a solution of the form
\begin{equation}
R(r)=r^{l}(1-r)e^{-\alpha r}\sum_{j=0}c_{j}r^{j}.  \label{eq:R_poly}
\end{equation}
It is not difficult to verify that the expansion coefficients $c_{j}$
satisfy the three-term recurrence relation
\begin{eqnarray}
c_{j+2} &=&A_{j}c_{j+1}+B_{j}c_{j},\;j=0,1,\ldots   \nonumber \\
A_{j} &=&\frac{2\alpha \left( j+l+2\right) -2\beta +j^{2}+j\left(
2l+5\right) +2\left( 2l+3\right) }{\left( j+2\right) \left( j+2l+3\right) },
\nonumber \\
B_{j} &=&2\frac{\beta -\alpha \left( j+l+2\right) }{\left( j+2\right) \left(
j+2l+3\right) },  \label{eq:TTRR_alpha}
\end{eqnarray}
if $E=-\alpha ^{2}/2$.

In order to obtain exact polynomial solutions we require that $c_{\nu }\neq 0
$ and $c_{\nu +1}=c_{\nu +2}=0$, $\nu =0,1,\ldots $. These conditions lead
to $B_{\nu }=0$ from which we obtain
\begin{equation}
\alpha =\frac{\beta }{l+\nu +2},\;E=-\frac{\beta ^{2}}{2(l+\nu +2)^{2}}.
\label{eq:alpha,E}
\end{equation}
Therefore

\begin{eqnarray}
A_{j} &=&\frac{2\beta \left( j-\nu \right) +\left( j^{2}+j\left( 2l+5\right)
+2\left( 2l+3\right) \right) \left( l+\nu +2\right) }{\left( j+2\right)
\left( j+2l+3\right) \left( l+\nu +2\right) },  \nonumber \\
B_{j} &=&\frac{2\beta \left( \nu -j\right) }{\left( j+2\right) \left(
j+2l+3\right) \left( l+\nu +2\right) }.  \label{eq:A_j,B_j_n}
\end{eqnarray}

The expression for $E$ in equation (\ref{eq:alpha,E}) does not give us the
spectrum of the problem. Note that $E=0$ when $\beta =0$ while the
Hamiltonian (\ref{eq:H_dim_beta}) tells us that we should obtain the
spectrum of the particle in a box of radius $r_{0}=1$ when $\beta =0$.
Besides, the polynomial solutions only provide negative eigenvalues while
all the eigenvalues of the model are positive for sufficiently small values
of $\beta $ as argued in section~\ref{sec:model}. Any smart reader may think
that it is not necessary to stress such an obvious fact but unfortunately
many researchers have misinterpreted the polynomial solutions of several
conditionally-solvable models as discussed elsewhere\cite{AF21, F21}.

Since $B_{\nu }=0$ the only remaining condition is $c_{\nu +1}=0$ from which
we obtain $\nu +1$ roots $\beta _{l}^{(\nu ,i)}$, $i=0,1,,\ldots ,\nu $,
that we conveniently arrange so that $\beta _{l}^{(\nu ,i+1)}>\beta
_{l}^{(\nu ,i)}$. Thus, the energies of the polynomial solutions should be
more properly written as
\begin{equation}
E_{l}^{(\nu ,i)}=-\frac{\left[ \beta _{l}^{(\nu ,i)}\right] ^{2}}{2(l+\nu
+2)^{2}}.  \label{eq:E^(n,i)_l}
\end{equation}
In the expressions above $\nu $ is the degree of the polynomial factor of
the exact solution (\ref{eq:R_poly}) and one can verify that $i$ is the
number of real zeros in the interval $0<r<1$. For this reason, $i$ (and not $%
\nu $) is the radial quantum number $n$. This fact was overlooked by many
researchers as discussed in the papers just mentioned\cite{AF21, F21}.

Since $i=n$ we conclude that $E_{nl}\left( 1,\beta _{l}^{(\nu ,n)}\right)
=E_{l}^{(\nu ,n)}<0$. The Hellmann-Feynman theorem\cite{G32,F39}
\begin{equation}
\frac{dE_{nl}(1,\beta )}{d\beta }=-\left\langle \frac{1}{r}\right\rangle
_{nl},  \label{eq:HFT}
\end{equation}
tells us that $E_{nl}$ decreases with $\beta $. Since $E_{nl}\left( 1,\beta
_{nl}^{c}\right) =0$ we conclude that $\beta _{l}^{(\nu ,n)}>\beta _{nl}^{c}$%
. Numerical results show that $\beta _{l}^{(\nu ,n)}$ decreases with $\nu $
as shown in Table~\ref{tab:betanu} for $\beta _{0}^{(\nu ,n)}$, $n=0,1,2,3$.
From these results and $\lim\limits_{\nu \rightarrow \infty }E_{l}^{(\nu
,n)}=0$ we may reasonably put forward the following

\begin{conj}
\label{c1} $\lim\limits_{\nu \rightarrow \infty }\beta _{l}^{(\nu ,n)}=\beta
_{nl}^{c}$
\end{conj}

In section~\-\ref{sec:RR} we will show numerical results that support this
conjecture. Of particular interest are the roots
\begin{equation}
\beta _{l}=\beta _{l}^{(0,0)}=(l+1)(l+2),\;E_{l}^{(0,0)}=-\frac{(l+1)^{2}}{2}%
,  \label{eq:beta_l}
\end{equation}
as shown below.

\section{Accurate numerical results}

\label{sec:RR}

One can obtain accurate numerical energies for the hydrogen atom in an
spherical box in several ways as shown in suitable reviews on the subject%
\cite{F01,SBC09a,SBC09b,S14}. Here, we resort to the RRM\cite{P68,SO96} that
provides increasingly accurate upper bounds to the exact eigenvalues\cite
{M33,F25a}.

For simplicity, we choose the non-orthogonal basis set
\begin{equation}
f_{il}(r)=r^{i+l}(1-r),\;i=0,1,\ldots .  \label{eq:basis}
\end{equation}
The RRM secular equations are well-known\cite{P68,SO96,F25a,F24} and we will
just outline them in what follows. In order to solve the radial equation (%
\ref{eq:radial}) we propose and ansatz of the form
\begin{equation}
\varphi (r)=\sum_{i=0}^{N-1}c_{i}f_{il}(r),
\end{equation}
and the RRM leads to the secular equation
\begin{equation}
\mathbf{Hc}=W\mathbf{Sc},
\end{equation}
where $\mathbf{H}$ and $\mathbf{S}$ are $N\times N$ matrices with elements $%
H_{ij}=\left\langle f_{il}\right| \mathcal{H}\left| f_{jl}\right\rangle $
and $S_{ij}=\left\langle f_{il}\right. \left| f_{jl}\right\rangle $,
respectively, and $\mathbf{c}$ is a $N\times 1$ column vector with elements $%
c_{i}$. The approximate eigenvalues $W_{nl}$, $n=0,1,\ldots ,N-1$, are roots
of the secular determinant $\left| \mathbf{H}-W\mathbf{S}\right| =0$. They
approach the exact eigenvalues $E_{nl}$ from above which facilitates the
estimation of the accuracy of the calculation. In the present case
\begin{equation}
\left\langle f\right. \left| g\right\rangle =\int_{0}^{1}f(r)g(r)r^{2}\,dr.
\end{equation}

Table~\ref{tab:EPB} shows some eigenvalues $E_{nl}^{PB}=E_{nl}(1,0)$. We
appreciate that there are several cases in which $E_{nl}^{PB}<E_{n^{\prime
}l^{\prime }}^{PB}$ when $n+l>n^{\prime }+l^{\prime }$. Consequently, we
expect that such eigenvalues $E_{nl}$ and $E_{n^{\prime }l^{\prime }}$
should cross at some nonzero value of $\beta $ because $E_{nl}^{H}>$ $%
E_{n^{\prime }l^{\prime }}^{H}$ as argued in section~\ref{sec:model}.

\begin{figure}[tbp]
\begin{center}
\includegraphics[width=9cm]{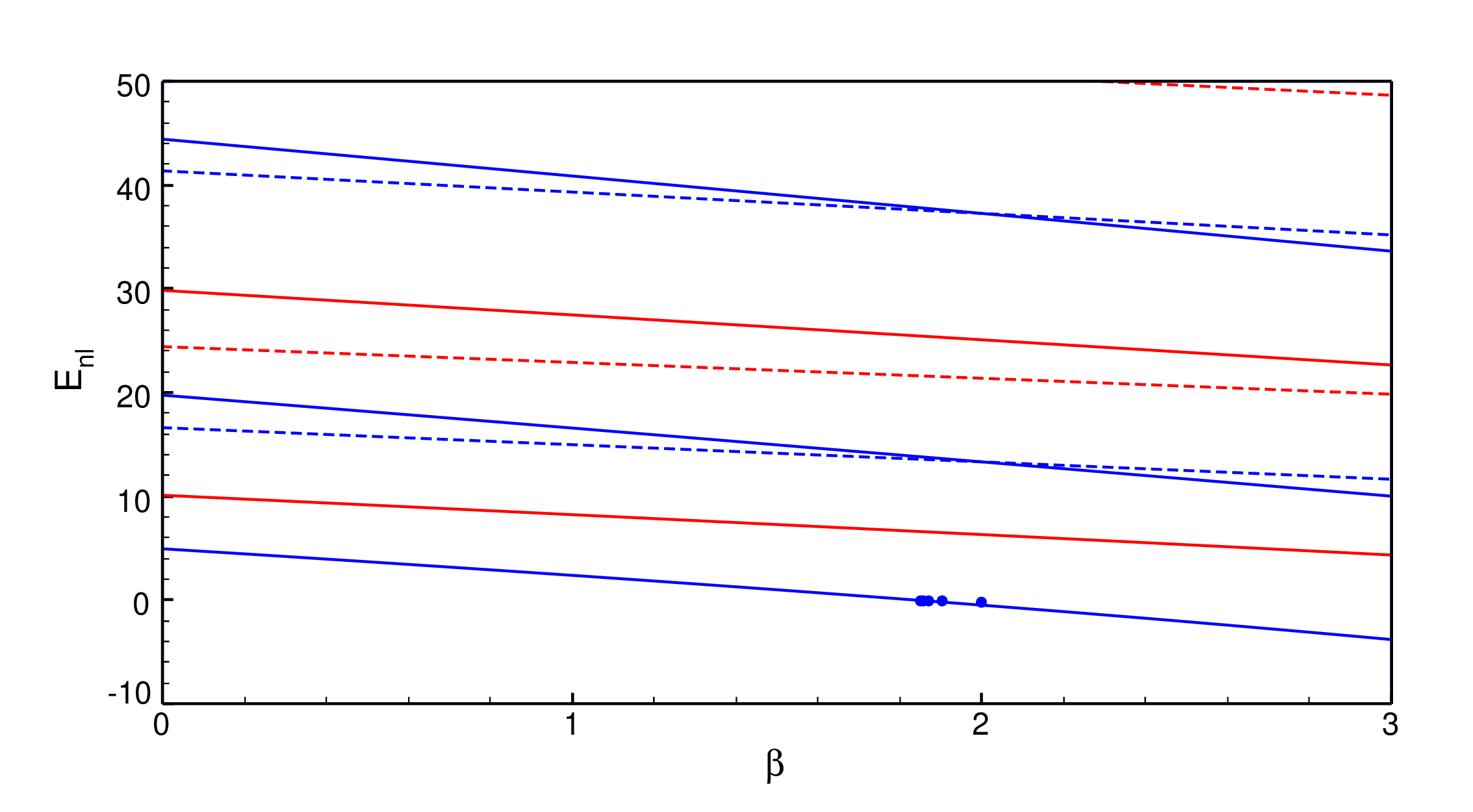}
\end{center}
\caption{Lowest eigenvalues with $l=0$ (blue solid lines), $l=1$ (red solid
lines), $l=2$ (blue dashed lines), $l=3$ (red dashed lines)}
\label{Fig:E(beta)}
\end{figure}

Figure~\ref{Fig:E(beta)} shows the lowest eigenvalues with $l=0,1,2,3$. We
appreciate the crossings at $\beta =\beta _{0}=2$ between $E_{10}$ and $%
E_{02}$ and also between $E_{20}$ and $E_{12}$. This fact suggests that the
values $\beta _{l}$ (\ref{eq:beta_l}) of the model parameter given by the
truncation condition are special. It is worth noting that the former
accidental degeneracy at $\beta =2$ appeared in an earlier paper\cite{K81}
(see also table~4 in page 140 in reference\cite{SBC09a}) but nobody paid
attention to it as far as we know. The blue points in figure~\ref
{Fig:E(beta)} are values of exact energies given by equation (\ref
{eq:E^(n,i)_l}) when $i=0$. Since the polynomial factors of such solutions
do not exhibit nodes, then they correspond to the ground state as the figure
already shows.

Table~\ref{tab:Crossings} shows several RRM eigenvalues calculated at $\beta
=\beta _{l}$, $l=0,1,2$. It is worth noting that the RRM yields the exact
eigenvalue $E_{0l}$ at $\beta =\beta _{l}$. From these results we draw the
following

\begin{conj}
\label{c2}  Pairs of eigenvalues $\left(
E_{n+1\,l},E_{n\,l+2}\right) $, $n=0,1,\ldots ,l=0,1,\ldots $\ cross at $%
\beta =\beta _{l}$
\end{conj}

At present we are unable to prove this conjecture rigorously.

\begin{figure}[tbp]
\begin{center}
\includegraphics[width=9cm]{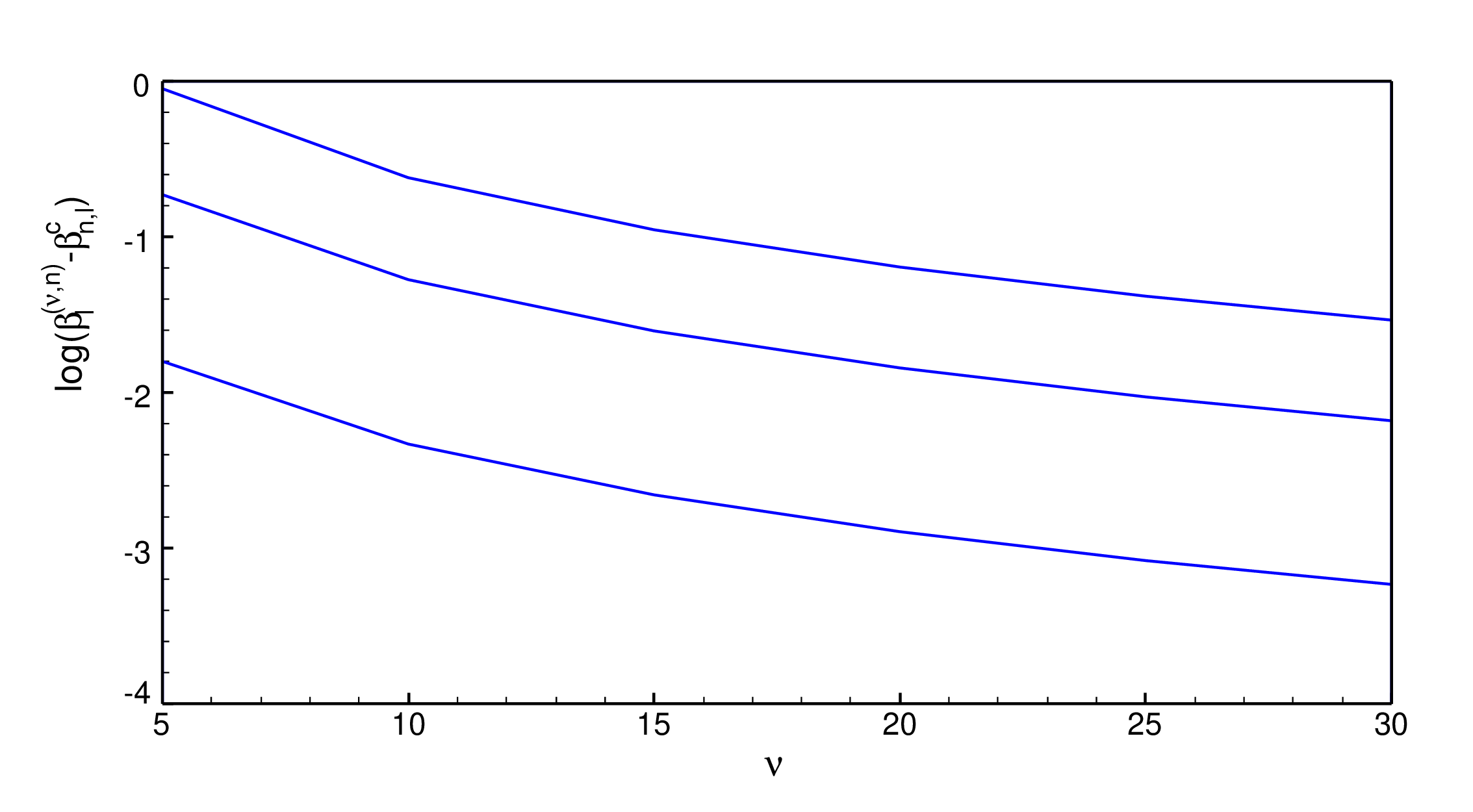}
\end{center}
\caption{$\log \left(\beta_l^{(\nu,n)}-\beta_{nl}^c \right)$}
\label{Fig:logconv}
\end{figure}

The RRM enables us to obtain the critical values of $\beta $ introduced in
section~\ref{sec:model}. We simply set $E=0$ in the secular equation and
solve for $\beta $. Table~\ref{tab:beta_crit} shows some critical values of $%
\beta $ for $l=0,1,2,3$. As discussed in section~\ref{sec:CS} the roots $%
\beta _{l}^{(\nu ,n)}$ approach to $\beta _{nl}^{c}$ from above when $\nu $
increases. Figure~\ref{Fig:logconv}, illustrates the rate of convergence.

\section{Conclusions}

\label{sec:conclusions}

In this paper we have shown several aspects of the well known hydrogen atom
inside a box with an impenetrable spherical wall that have passed unnoticed,
as far as we know. In the first place, a suitable scaling of the Hamiltonian
operator is extremely useful for the derivation of several general
properties of the eigenvalues. In the second place, the radial part of the
Schr\"{o}dinger equation is conditionally solvable. In the third place,
there are most interesting seemingly accidental degeneracies that take place
at particular values of the box radius that are determined by a truncation
condition. In the fourth place, some of the roots given by the truncation
condition appear to converge towards the critical values of the model
parameter. At present we cannot prove the two latter results rigorously and
have, therefore, presented them as conjectures. In this analysis the RRM
proved to be most useful.

\begin{table}[tbp]
\caption{$\beta^{(\nu,n)}_0$ for increasing values of $\nu$}
\label{tab:betanu}%
\begin{tabular}{rD{.}{.}{10}D{.}{.}{10}D{.}{.}{10}D{.}{.}{10}}
\multicolumn{1}{c}{$\nu$} & \multicolumn{1}{c}{$n=0$} &
\multicolumn{1}{c}{$n=1$} & \multicolumn{1}{c}{$n=2$} &
\multicolumn{1}{c}{$n=3$} \\

  5 & 1.846838425 & 6.287049333 & 13.56824532 & 24.21585798  \\
 10 & 1.839160212 & 6.196784392 & 13.13767165 & 22.79428563  \\
 15 & 1.837192594 & 6.174300067 & 13.03554186 & 22.48225348  \\
 20 & 1.836407529 & 6.165399473 & 12.99562834 & 22.36253197  \\
 25 & 1.836016970 & 6.160986334 & 12.97594386 & 22.30392905  \\
 30 & 1.835794835 & 6.158480669 & 12.96479836 & 22.27087450  \\

\end{tabular}
\end{table}

\begin{table}[tbp]
\caption{Some eigenvalues for $\beta =0$}
\label{tab:EPB}
\begin{center}
\begin{tabular}{cD{.}{.}{10}}

\multicolumn{1}{c}{$(n,l)$}  & \multicolumn{1}{c}{$E_{nl}^{PB}$}
\\ \hline \hline
(0,0)&  4.934802200 \\\hline
(0,1)&  10.09536427 \\
(1,0)&  19.73920880 \\\hline
(0,2)&  16.60873095 \\
(1,1)&  29.83975797 \\
(2,0)&  44.41321980 \\\hline
(0,3)&  24.41559682 \\
(1,2)&  41.35961555 \\
(2,1)&  59.44993458 \\
(3,0)&  78.95683520 \\\hline
(0,4)&  33.47715596 \\
(1,3)&  54.25817941 \\
(2,2)&  75.92743708 \\
(3,1)&  98.92890559 \\
(4,0)&  123.3700550 \\\hline
(0,5)&  43.76561012 \\
(1,4)&  68.50242574 \\
(2,3)&  93.81791915 \\
(3,2)&  120.3514532 \\
(4,1)&  148.2772060 \\
(5,0)&  177.6528792 \\\hline
(0,6)&  55.25985415 \\
(1,5)&  84.06545236 \\
(2,4)&  113.0957572 \\
(3,3)&  143.2044787 \\
(4,2)&  174.6400399 \\
(5,1)&  207.4949921 \\

\end{tabular}
\end{center}
\end{table}

\begin{table}[tbp]
\caption{$E_{nl}(1,\beta_l)$ for some values of $n$ and $l$}
\label{tab:Crossings}%
\begin{tabular}{cD{.}{.}{10}D{.}{.}{10}D{.}{.}{10}D{.}{.}{10}}

l   & \multicolumn{1}{c}{$n=0$}   &   \multicolumn{1}{c}{$n=1$}
&     \multicolumn{1}{c}{$n=2$}
&     \multicolumn{1}{c}{$n=3$} \\
                  \multicolumn{5}{c}{$\beta_0=2$}            \\
0 & -0.5   &       13.31003662 & 37.25660174 & 71.26437398  \\
2 &  13.31003662 & 37.25660174 & 71.26437398 & 115.2540228 \\

                \multicolumn{5}{c}{$\beta_1=6$}              \\
1 & -2    &       15.17434035 & 42.95936431 & 81.04494034  \\
3 & 15.17434035 & 42.95936431 & 81.04494034 & 129.2643219  \\
                                                             \\
               \multicolumn{5}{c}{$\beta_2=12$}               \\
2 & -4.5   &      15.84159512 & 47.2388141  & 89.18513747 \\
4 & 15.84159512 & 47.2388141 & 89.18513747 & 141.4317571  \\
\end{tabular}
\end{table}

\begin{table}[tbp]
\caption{Some critical values of $\beta$}
\label{tab:beta_crit}%
\begin{tabular}{cD{.}{.}{10}D{.}{.}{10}D{.}{.}{10}D{.}{.}{10}}

l &     \multicolumn{1}{c}{$\beta_{0,l}^c$}          &
\multicolumn{1}{c}{$\beta_{1,l}^c$}
&      \multicolumn{1}{c}{$\beta_{2,l}^c$}        &  \multicolumn{1}{c}{$\beta_{3,l}^c$}             \\
0 &   1.835246330 &   6.152307040 &  12.93743173 &  22.19009585  \\
1 &   5.088308227 &  11.90969656  &  21.17443122 &  32.90010678  \\
2 &   9.617366041 &  19.03014419  &  30.81193326 &  45.03068523  \\
3 &  15.36345002  &  27.45875083  &  41.80446073 &  58.54453721  \\
\end{tabular}
\end{table}

\end{document}